\documentclass[twocolumn, aps, prb 5p,sort&compress]{revtex4-1}
 \usepackage{amsmath,amssymb}
 \usepackage{stmaryrd}
 \usepackage{latexsym}
 \usepackage{CJK}
 \usepackage{graphicx}
 \usepackage{indentfirst}
 \usepackage{listings}
 \usepackage{xcolor}
 \usepackage{booktabs}
 \usepackage{stmaryrd}
 \usepackage{multirow}
 \usepackage{verbatim}
 \usepackage{makecell}

\begin{document}

   \title{Grain Boundary Shear Coupling is Not a Grain Boundary Property}
   \author{Kongtao Chen$^1$}
   \author{Jian Han$^1$}
   \author{Spencer L. Thomas$^1$}
   \author{David J. Srolovitz$^{1,2,3}$}
%\cortext[mycorrespondingauthor]{Corresponding author. Tel.: +1 (215) 898-6924}

\affiliation{$^1$Department of Materials Science and Engineering, University of Pennsylvania, Philadelphia, Pennsylvania 19104, USA}
\affiliation{$^2$Department of Mechanical Engineering and Applied Mechanics, University of Pennsylvania, Philadelphia, Pennsylvania 19104, USA}
\affiliation{$^3$Department of Materials Science and Engineering, City University of Hong Kong, Hong Kong SAR, PRC}

   \date{\today}

   \begin{abstract}
 %{\color{blue}{Shear coupling is crucial for grain boundary(GB)'s kinetics but temperature, stress, and chemical potential jump dependence of coupling factor is unclear.  So we introduce the constant stress simulation method for shear coupling and compare it with conventional constant strain rate one.  And then we study temperature and driving force dependence of coupling factor under constant stress and chemical potential jump by molecular dynamics (MD).  We found that the coupling factor \(\beta\) depends on not only temperature, but also on the type and magnitude of driving force, quantitatively consistent with our multi-mode analytical model. It indicates that GB moves by nucleating different types of disconnections according to temperature and driving force.}}

Shear coupling implies that all grain boundary (GB) migration necessarily creates mechanical stresses/strains and is a key component to the evolution of all polycrystalline microstructures.
We present MD simulation data and theoretical analyses that demonstrate the GB shear coupling is not an intrinsic GB property, but rather strongly depends on the type and magnitude of the driving force for migration and temperature.
We resolve this apparent paradox by proposing a microscopic theory for GB migration that is based upon a statistical ensemble of line defects (disconnections) that are constrained to lie in the GB.
Comparison with the MD results for several GBs provides quantitative validation of the theory as a function of stress, chemical potential jump and temperature.

   \end{abstract}

   \maketitle

   \section*{Introduction}
Grain boundary (GB) motion is central to a wide range of microstructure evolution processes, including normal grain growth, abnormal grain growth, primary recrystallization, and sintering.
These processes may be driven by different factors; e.g., stress,  injection of defects from within the grains, capillarity (surface tension), and differences in defect densities or elastic energy (across the GB).
Shear coupling refers to the motion of GBs driven by shear across the GB plane or, equivalently, the displacement of one grain relative to the other during GB migration.
While shear coupling was first observed over 60 years ago \cite{li1953stress,bainbridge1954recent,biscondi1968intercrystalline}, interest in this topic has grown considerably in the past 30 years.
Shear coupling has been reported in both metals (e.g., Al \cite{biscondi1968intercrystalline, fukutomi1991sliding, winning2001stress, winning2002mechanisms, winning2005transition}, Zn \cite{li1953stress, bainbridge1954recent}) and ceramics (e.g., cubic zirconia \cite{yoshida2004high}).
Grain boundary sliding is, in some sense, the absence of shear coupling (shear across the grain boundary produces no migration).
Grain boundary sliding has been observed in a wide range of polycrystalline systems (e.g., see [\onlinecite{sheikh2003stimulation}]).
These experimental observations of shear coupling and grain boundary sliding have been reproduced in a wide-range of atomic-scale simulations  (e.g., see [\onlinecite{molteni1996first, molteni1997sliding,hamilton2002first, chen1992finite, shiga2004stress, chandra1999atomistic, haslam2003stress, sansoz2005mechanical, Cahn2006, thomas2017reconciling}]).

The shear coupling factor  \(\beta=v_{||}/v_{\perp}\) is the ratio of the shear rate across the GB (\(v_{||}\)) to the normal (migration) velocity of the GB  (\(v_{\perp}\)).
Interestingly, molecular dynamics (MD) simulations of shear coupling under a fixed shear strain rate~\cite{Cahn2006} and those performed based on a synthetic driving force~\cite{homer2013phenomenology} have shown very different values of \(\beta\) for the same grain boundary.
(A synthetic driving force is a simulation method for producing a jump in chemical potential across a GB; physically, such jumps may result from the capillarity/Gibbs-Thompson effect, differences in defect densities, and differences in elastic strain energy differences associated with elastic anisotropy.)
Additionally, both simulations~\cite{Cahn2006} and experiments~\cite{gorkaya2010experimental} demonstrate that \(\beta\) is often temperature dependent; i.e., in some cases, \(\beta\rightarrow\infty\) at high temperature - GB sliding.
In this paper, we examine how \(\beta\) varies with the type of driving force, the magnitude of the driving force, and temperature.
In particular, we perform MD simulation of GB motion driven by an applied shear stress, (the more widely used) an applied shear strain rate, and a jump in chemical potential across the GB for different driving force magnitudes and temperature for several crystallographically distinct GBs.
In short, we find that \(\beta\) varies with all three of these factors (type and magnitude of the driving force and temperature).
We propose an approach to understand these observations based upon the microscopic mechanism of GB migration, i.e., disconnection motion  \cite{han2018grain}, as well as the competition between different disconnection modes \cite{thomas2017reconciling}.
Based on this approach, we make quantitative predictions of how shear coupling varies with both the type and magnitude of the driving force and with temperature and validate these predictions against our MD results.

\section*{Mechanically-Driven Shear Coupling}
The mechanical deformation of a polycrystal, whether under stress or strain-control, results in non-uniform stresses and strains within the sample.
Strain-controlled and stress-controlled loading can lead to very different deformation behavior.
In computer modeling and theoretical treatments of the reaction of grain boundaries to mechanical deformation, the loading is most commonly applied under fixed strain-rate conditions~\cite{Cahn2006}.
Bicrystal shear coupling experiments are most commonly performed under fixed stress~\cite{rupert2009experimental}.
Here, we investigate the difference in shear coupling associated with fixed stress and fixed strain rate loading.
Constant stress simulations are more easily analyzed in a statistical mechanics framework (see below) than their constant strain rate counterparts.
We perform MD simulations of several symmetric tilt GBs in copper using periodic boundary conditions.
In particular, we examine the \(\Sigma5[001](310)\),  \(\Sigma13[001](510)\),  \(\Sigma37[001](750)\) and the \(\Sigma7[111](12\bar{3})\) symmetric tilt GBs.
The simulation cell is periodic in all directions, where a pair of nominally flat, parallel GBs have normal \(y\) and the tilt axis is parallel to \(z\).
A constant shear stress \(\sigma_{xy}\) is applied, while the other elastic fields satisfy: \(\sigma_{yz}=\sigma_{yy}=0\) and \(\epsilon_{xx}=\epsilon_{zz}=\epsilon_{xz}=0\).
We also perform MD simulation in exactly the same simulation cells at fixed shear strain rate \(\dot{\epsilon}_{xy}\) together with \(\sigma_{yz}=\sigma_{yy}=0\) and \(\epsilon_{xx}=\epsilon_{zz}=\epsilon_{xz}=0\).  (See Methods and  the Supplementary Information for more detail).

\begin{figure}[!hb]
      \centering
      \includegraphics[scale=0.3]{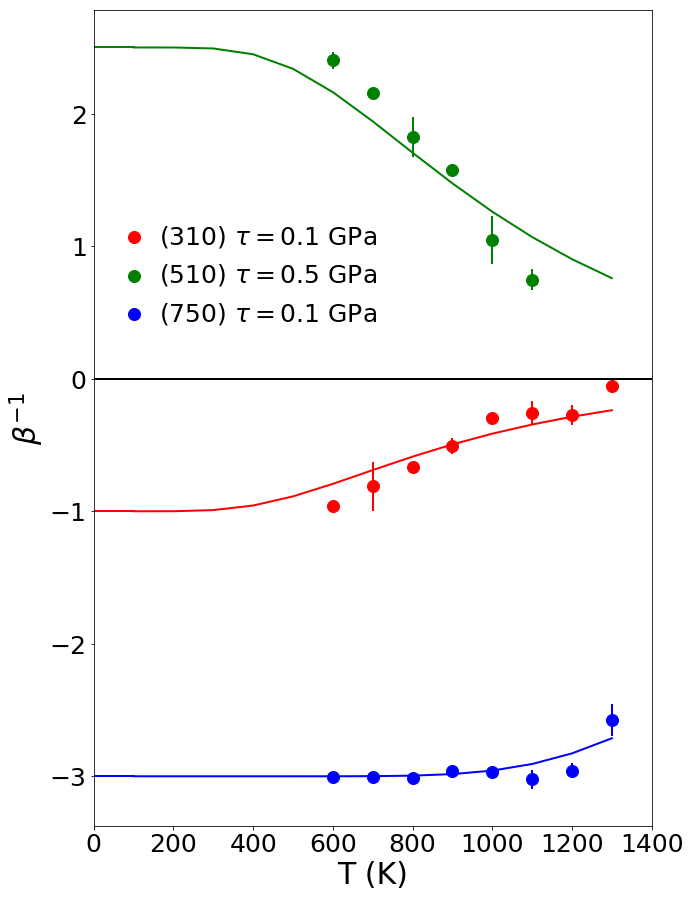}
      \caption{Temperature dependence of the inverse shear coupling factor  \(\beta^{-1}\) for  shear stress-driven  migration of three [001] symmetric tilt grain boundaries,  \(\Sigma5(310)\), \(\Sigma13(510)\), and \(\Sigma37(750)\).  The circles indicate the mean value of  \(\beta^{-1}\) for two GBs (the individual values of \(\beta^{-1}\) are the top and bottom of the error bars which are not visible when their difference is smaller than the size of the circle). The continuous curves are  from the fits to Eq. \eqref{betaMulModel} and \eqref{barrier} for each GB.}\label{BTP}
\end{figure}

Figure~\ref{BTP} shows the inverse shear coupling factor  \(\beta^{-1}\) versus temperature for shear stress  (\(\sigma_{xy}=\tau\)) driven migration of three symmetric tilt grain boundaries.
In all cases, \(|\beta^{-1}|\) decreases with increasing temperature.
While this decrease is particular evident for the \(\Sigma5(310)\) and \(\Sigma13(510)\) GBs, the \(\Sigma37(750)\) is nearly temperature independent until very near the melting point  (\(T_\text{m}=1320\) K for this interatomic potential~\cite{Mishin2011}).
This tendency is consistent with earlier simulations in fixed strain rate-driven shear coupling (e.g., see [\onlinecite{Cahn2006}]).
The decrease in  \(|\beta^{-1}|\) with increasing temperature is consistent with the widely known increase in the grain boundary sliding rate with increasing temperature~\cite{Cahn2006}.

   \begin{figure}[!htb]
      \centering
      \includegraphics[scale=0.3]{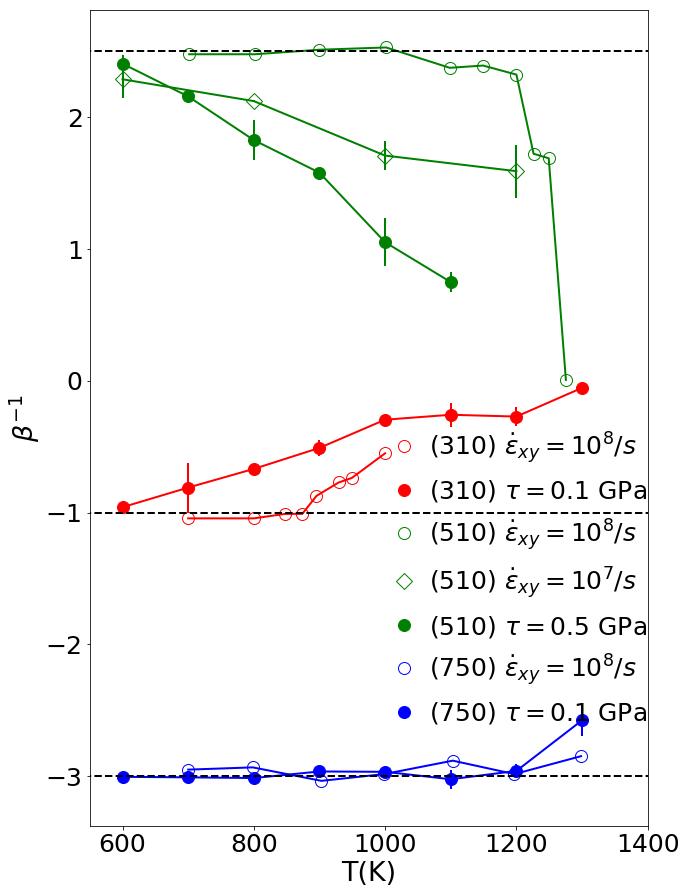}
      \caption{The temperature dependence of the inverse coupling factor \(\beta^{-1}\) is different under constant stress and strain rate. Constant stress MD data and diamond symbol constant train rate data are averaged over 2 GBs. The open circles correspond to the constant strain rate (fixed end boundary condition) MD data from ~[\onlinecite{Cahn2006}]. The open diamond symbols show constant strain rate simulation results under periodic boundary conditions. The horizontal black dashed lines indicate the values of \(\beta^{-1}\) for \(T\to 0\).}\label{BTC}
    \end{figure}

Figure~\ref{BTC} shows the inverse shear coupling factor \(\beta^{-1}\) versus temperature for fixed shear strain rate  \(\dot{\epsilon}_{xy}\) driven migration of the same three symmetric tilt grain boundaries, where the  open circles  are from~[\onlinecite{Cahn2006}] and the diamonds are from the present work.
Since the simulation cell size was not given explicitly in  [\onlinecite{Cahn2006}], we estimate the shear rate employed based on the bicrystal size  (7 nm - in the direction normal to the GB plane) from the figures  \cite{Cahn2006} and the given shear velocity; i.e.,  \(\dot{\epsilon}_{xy} \sim 10^8/s\). Our simulation cell was approximately ten times larger and the shear strain rate is \(\sim\) an order of magnitude smaller (our simulation cell also was periodic in the direction normal to the GB plane and contained two GBs.
The two data sets for the same (510) GB  show similar tendencies although the lower strain rate data exhibits smaller values of \(\beta^{-1}\) than those at larger strain rate.

    \begin{figure}[!htb]
      \centering
      \includegraphics[scale=0.3]{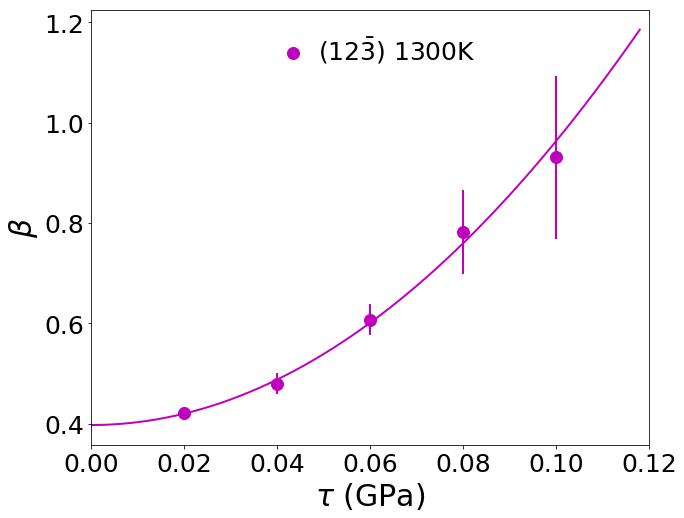}
      \caption{Coupling factor versus shear stress for the \(\Sigma7[111](12\bar{3})\) symmetric tilt GB at 1300 K.  The data points are represent the mean for 2 GBs. The continuous curve is the best fit parabola to these data, as suggested by Eq. \eqref{fitBetaStress}.}\label{BP2}
    \end{figure}

Figure~\ref{BTC} also shows a comparison between the shear coupling factors obtained under constant stress and constant strain rate conditions.
The stress-driven migration simulations were performed using the same simulation cell size and periodic boundary conditions as in our fixed strain rate simulations.
These data show that the absolute value of the inverse coupling factor \(|\beta^{-1}|\) is larger for the fixed strain rate simulations than for the fixed stress simulations for the (310) and (510) GBs (since \(\beta^{-1}\) for the (750) GB is nearly temperature-independent, no conclusions can be drawn from this GB).
In addition, in almost every case, the variation of the slope of the absolute value of the inverse coupling factor  with temperature (\(\partial|\beta^{-1}|/\partial T\)) is larger for the fixed stress simulations than for the fixed strain rate simulations.
Since \(\beta\) is a function of temperature, constant strain rate  implies that the shear stress is a function of temperature.
And, correspondingly, constant stress  implies that the strain rate is a function of temperature.
The fact that constant stress and constant strain rate loading give different results is not surprising in light of the differences in stress-strain response under different loading conditions during plastic deformation.
We present a quantitative explanation for this observation below.

It is widely accepted that at low temperature, the coupling factor is controlled by GB geometry~\cite{Cahn2006} and is independent of mechanical load;  this is consistent with our simulation results (not shown).
However, Fig. \ref{BP2} shows that the coupling factor \(\beta\) is a function of shear stress at high temperature.

A quantitative model describing the dependence of the shear coupling factor on both temperature and the magnitude of the mechanical loading is presented below.

\section*{Chemical Potential Jump-Driven Shear Coupling}
Shear coupling may also occur during GB migration when it is induced by non-mechanical driving forces.
Following Janssens et al.~\cite{janssens2006computing}, we simulate GB migration driven by a jump in chemical potential across the GB, \(\psi\), under periodic boundary conditions where the entire simulation cell may shear to maintain zero-net shear stress.
Additional MD simulations are performed in which GB migration is driven by a jump in chemical potential
\(\psi\) (\(= \psi^+-\psi^-\), where \(\psi^\pm\) indicate the chemical potential above/below the GB)
 across the GB while keeping \(\sigma_{yz}=\sigma_{yy}=0\) and \(\epsilon_{xx}=\epsilon_{zz}=\epsilon_{xz}=0\).

\begin{figure}[!htb]
      \centering
      \includegraphics[scale=0.3]{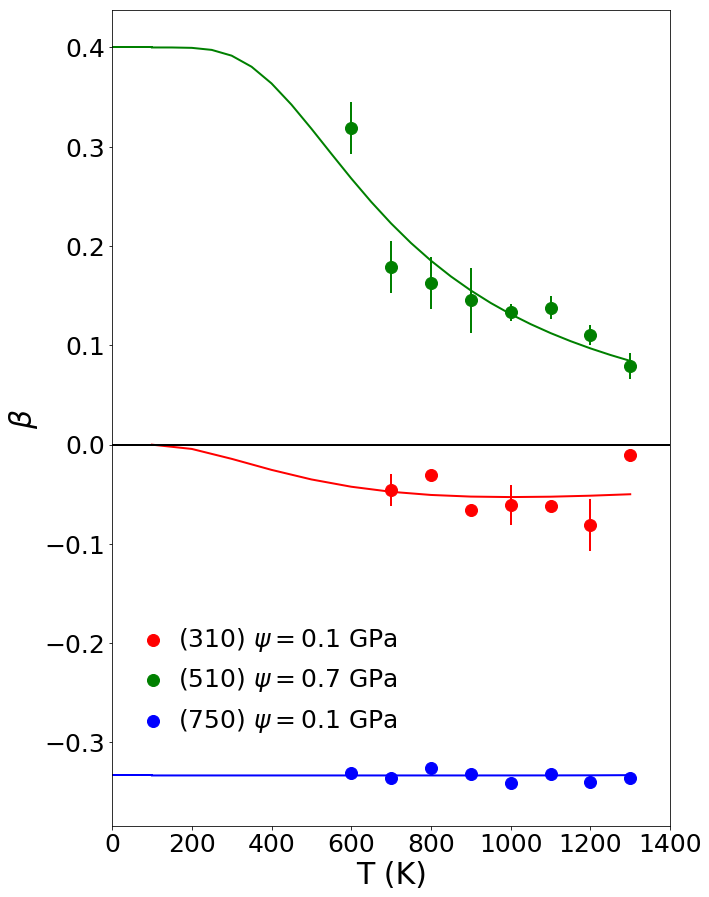}
      \caption{Temperature dependence of the coupling factor \(\beta\) for three GBs driven by a constant chemical potential jump under zero net shear stress conditions.  The continuous curves are   fits of the data to Eqs. \eqref{betaMulModel} and \eqref{barrier} for each GB.}\label{BTE}
\end{figure}

\begin{figure}[!hb]
      \centering
      \includegraphics[scale=0.3]{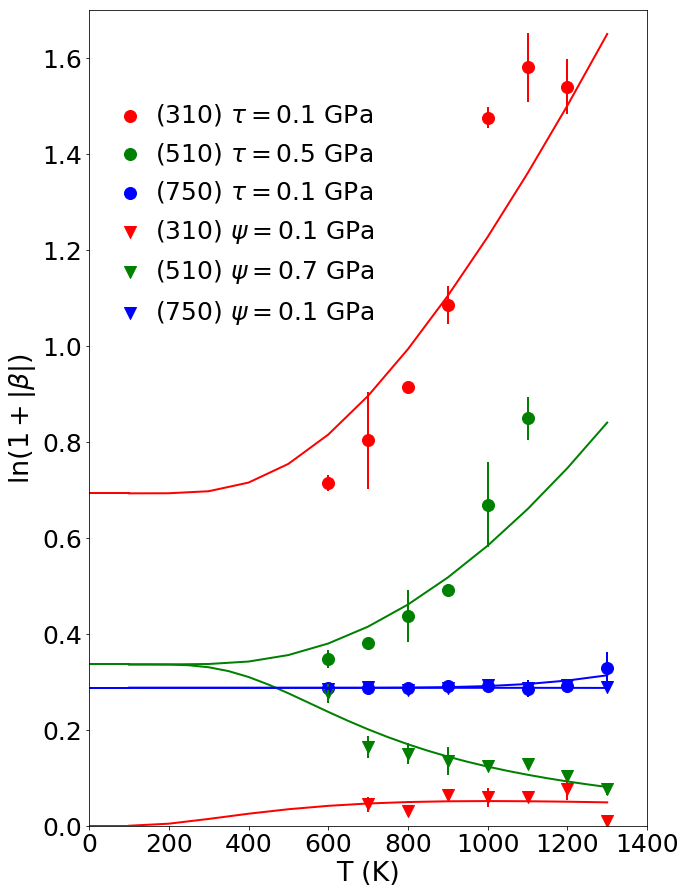}
      \caption{Temperature dependence of the coupling factor \(\beta\) for constant stress and chemical potential jump driving forces. The data are plotted on a logarithmic scale to fit the data meaningfully on one plot.
The continuous curves are fits to the constant stress and chemical potential jump data as per Eqs. \eqref{betaMulModel} and \eqref{barrier} for each GB.
           Curves are from fitting A and B from Eqs. (\ref{betaMulModel}) and (\ref{barrier}) for each GB.}\label{BTP2}
    \end{figure}

Figure \ref{BTE} shows the temperature dependence of the shear coupling factor for this driving force for the same three GBs discussed above.
For the  \(\Sigma 13(510)\) GB, \(\beta\) decreases with increasing temperature, while for the  \(\Sigma 5(310)\)  and \(\Sigma 37(750)\) GBs, the coupling factor is nearly temperature independent.
This is in stark contrast with \(\beta(T)\) for the mechanically-driven GB migration results in Fig. \ref{BTP}, especially for the \(\Sigma 5(310)\) and  \(\Sigma13(510)\) GB cases (we provide a direct comparison in Fig. \ref{BTP2}).

\begin{figure}[!htb]
      \centering
      \includegraphics[scale=0.3]{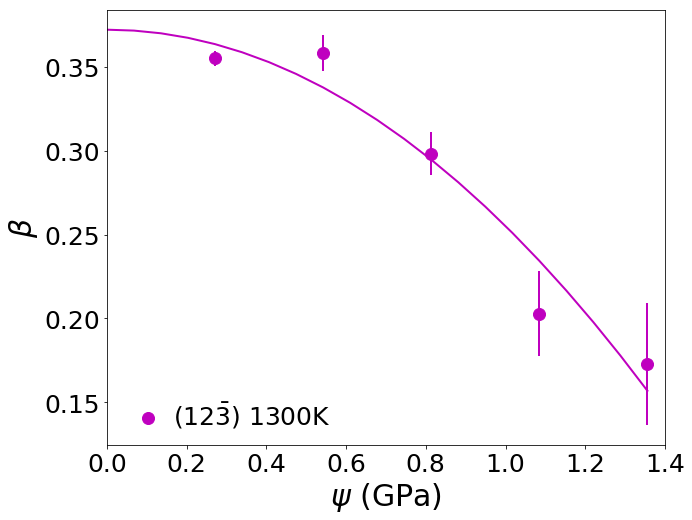}
      \caption{Coupling factor versus chemical potential jump for the \(\Sigma7[111](12\bar{3})\) symmetric tilt GB at 1300 K.  The data points  represent the mean for 2 GBs. The continuous curve is the best fit parabola to these data, as suggested per Eq. \eqref{fitBetaEnergy}.} \label{BE}
    \end{figure}

It is widely accepted that at low temperature, the coupling factor is controlled by GB geometry~\cite{Cahn2006} and is independent of chemical potential jump (this is consistent with our simulation results, not shown here).
However, Fig. \ref{BE} shows that the coupling factor \(\beta\) for the \(\Sigma7[111](12\bar{3})\) symmetric tilt GB is a function of the magnitude of the chemical potential jump at high temperature (this effect is particularly striking for this GB).
While larger driving force leads to larger magnitude coupling factors \(|\beta|\) under mechanical driving forces (see Fig.~\ref{BP2}), larger chemical potential jump-driving forces lead to smaller magnitude coupling factors \(|\beta|\).
A quantitative model describing the dependence of the shear coupling factor on both temperature and the magnitude of the chemical potential jump is presented below.

Figure \ref{BTP2} shows a direct comparison of the shear coupling factors for different types of driving forces; i.e., shear stress and chemical potential jump.
(We plot this on a logarithmic-scale, \(\ln(1+|\beta|)\) vs. T, to fit all of these data on one plot.)
For the \(\Sigma 5(310)\) GB, the stress and chemical potential jump coupling factor data are very different at both low and high temperature; this difference grows with increasing temperature.
For the \(\Sigma 13(510)\) GB, the values of \(\beta\) for the two types of driving forces are the same at low temperature but diverge with at higher temperature.
On the other hand, for the \(\Sigma 37(750)\) GB, the values of \(\beta\) for the two types of driving forces are nearly the same and temperature independent.
We present a microscopic mechanism-based analysis for the mode selection below.% in Table \ref{tab:parameter1}.

\section*{Statistical Disconnection Model}
Disconnections are line defects within an interface that are both dislocations and steps, characterized by a Burgers vector \(\mathbf{b}\) and step height \(h\), respectively.
For a given GB, permissible combinations (modes) of \((\mathbf{b},h)\) are completely determined by the bicrystallography \cite{han2018grain}.
While pure step modes (\(\mathbf{b}=\mathbf{0},h\neq0\)) and pure dislocation modes (\(\mathbf{b}\neq\mathbf{0},h=0\)) may exist, these  never correspond to both small \(b=|\mathbf{b}|\) and \(|h|\).
GBs migrate through the formation and migration of disconnections.
Therefore, grain boundary migration (resulting from step motion along the GB) and lateral grain translation (motion of one grain relative to the other across the GB, resulting from dislocation migration along the GB) are coupled (e.g., see \cite{Cahn2006, han2018grain}).
Disconnection migration may be driven by either through a jump in chemical potential across the GB or by a mechanical stress.
Shear stresses drive disconnection migration in much the same way that they drive the motion of lattice dislocations (as described by the Peach-Koehler equation).
Chemical potential jumps drive disconnection  motion through the motion of atoms across the GB (at steps) from the low to high chemical potential grains.

Since shear stress \(\tau\) couples (is conjugate) to \(\mathbf{b}\) and chemical potential jump \(\psi\) couples (is conjugate) to \(h\), the nucleation barrier for a pair of disconnections of mode \((\mathbf{b},h)\) depends on both. Following the detailed discussion of disconnection nucleation in [\onlinecite{han2018grain}]  (for the case of a straight dislocation dipole in an  isotropic, periodic system), we write the disconnection nucleation barrier as
\begin{equation}\label{qMulModel}
    q=(Ab^{2}+B|h|)L-bS\tau+hS\psi,
\end{equation}
where \(A=-2G\left[(1-\nu\cos^{2}\alpha)/4\pi(1-\nu)\right]\ln\left[\sin{(\pi r/w)}\right]\) and \(B=2\gamma\).
\(b\) is the magnitude of the Burgers vector that is conjugate to shear stress \(\tau\),  \(\gamma\) is the GB energy (per unit area), \(G\) and \(\nu\) are the shear modulus and Poisson's ratio, \(\alpha\) is the angle between the Burgers vector and the disconnection line direction, and \(r\) is the disconnection core size.
In the periodic unit cell employed in the MD simulations, \(L\) is the cell dimension parallel to the nominally straight disconnection lines, \(w\) is the cell dimension in the direction orthogonal to the disconnection line, and \(S=Lw\) is the nominal GB area.
\(A\)  describes the energy required to form a pair of dislocations and separate them to a distance of half the periodic unit cell (\(w/2\)) \cite{hirth1982theory} and
 \(B\) describes the energy required to form a pair of steps~\cite{han2018grain}.

Equation \eqref{qMulModel} suggests that mode selection depends on both the magnitude AND type of driving force (stress or chemical potential jump).
Large stresses favor modes of large \(|\mathbf{b}|\) and small \(|h|\), while large chemical potential jumps favor modes of small \(|\mathbf{b}|\) and large \(h\) (especially the pure step mode with \(\mathbf{b}=0\)), resulting in different coupling factors \(\beta\) in these cases;
the larger the driving force, the stronger this effect.

Since the disconnection nucleation barrier in Eq. \eqref{qMulModel} depends on \(\mathbf{b}\) and \(h\), we should expect that different disconnection modes will have different nucleation rates.
%Equation \eqref{qMulModel} also suggests that pure step mode plays different roles under stress and chemical potential jump. Under pure stress (\(\psi=0\)), pure step mode has no contribution to the summations in Eq. \eqref{qMulModel} since it nucleates evenly in either direction of GB, while it may affect \(\beta\) under chemical potential jump. Therefore, when a GB's migration is dominated by pure step mode (\(\beta\rightarrow0\)) under pure chemical potential jump, GB \((310)\) in Fig. \ref{BTP2} for example, \(\beta\) will differ a lot when it's driven by pure stress.
This effect may be captured via Boltzmann statistics~\cite{thomas2017reconciling}.
In this way, we describe the effective shear coupling factor by weighting the coupling factors associated with  disconnection modes \(i\) (\(\beta_i=b_i/h_i\)) by their Boltzmann factors
\begin{equation}\label{betaMulModel}
\begin{split}
    \beta &=\frac{\sum_{i}b_{i}e^{-\frac{Q_{i}}{k_{B}T}}\sinh{\frac{(b_{i}\tau-h_{i}\psi)S}{k_{B}T}}}{\sum_{i}h_{i}e^{-\frac{Q_{i}}{k_{B}T}}\sinh{\frac{(b_{i}\tau-h_{i}\psi)S}{k_{B}T}}}\\
    &=\frac{\sum_{i}b_{i}e^{-\frac{Q_{i}}{k_{B}T}}(b_{i}\tau-h_{i}\psi)}{\sum_{i}h_{i}e^{-\frac{Q_{i}}{k_{B}T}}(b_{i}\tau-h_{i}\psi)}+\mathcal{O}(\frac{(b\tau-h\psi)S}{k_{B}T})^2,
\end{split}
\end{equation}
where the summation is over all crystallographically possible disconnection modes,
\(k_{B} T\) is the thermal energy and
%Boltzmann constant, \(T\) is temperature, \(\tau\) is the shear stress and , and \(\psi\) are the coupling factor, Boltzmann constant, temperature, shear stress,  and chemical potential jump across GB, respectively.
\(Q_i\) is intrinsic disconnection nucleation barrier  for the \(i^\text{th}\) disconnection mode (i.e., \(Q_i=q\) in the absence of a driving force)
\begin{equation}\label{barrier}
        \frac{Q_i}{L}=Ab_i^{2}+B|h_i|.
\end{equation}
At low temperature, only the disconnection mode with the lowest barrier is activated, while at high temperature, many modes are activated, resulting in  \(\beta\) being a function of temperature.

\subsection*{Temperature and driving force type}

Before comparing the disconnection model prediction of \(\beta\) with the MD simulation results, we note that the expressions for \(A\) and \(B\) following Eq.~\eqref{qMulModel} represent continuum model descriptions of fundamentally atomic-level and bonding-dependent quantities (related to disconnection core structures).
As such, we treat \(A\) and \(B\) as parameters to be determined by fitting to the simulation data and defer the assessment of how well the analytical expressions for \(A\) and \(B\) work.
We perform nonlinear fits of Eqs. \eqref{betaMulModel} and \eqref{barrier} to the \(\beta\) data in Figs. \ref{BTP} (stress driving force) and \ref{BTE} (chemical potential driving force).
To do this, we consider all of the disconnection modes (in practice only the lowest few modes are important) for the \(\Sigma5[001](310)\),  \(\Sigma13[001](510)\),  and \(\Sigma37[001](750)\) symmetric tilt GBs (see [\onlinecite{han2018grain}]  for a description of how to enumerate all possible disconnection modes).
The results of this fitting procedure are shown as the continuous curves in Figs. \ref{BTP} and \ref{BTE} and in Table \ref{tab:parameter1}.
Overall, we see that Eqs. \eqref{betaMulModel} and \eqref{barrier} are in good agreement with the MD results for both driving forces as a function of temperature.
The predicted temperature dependence is especially remarkable giving the simplicity of the Boltzmann weighting of the different disconnection modes; even better agreement should be possible with inclusion of correlation effects, e.g., through the use of kinetic Monte Carlo approaches.

    \begin{table}[htbp]
    \setlength\extrarowheight{5pt}
    \caption{Fitting parameters $A$ and $B$ in Eq.~\eqref{barrier} for the data in Figs. \ref{BTP} and \ref{BTE}. $\gamma$ is the GB energy for this potential at 0 K~\cite{Mishin2011}. $b_i$, $h_i$, and $\beta_i$ are the Burgers vector, step height, and coupling factor of disconnection mode with the $i^\text{th}$ lowest barrier (see Eq.~\eqref{barrier} and the analytical expressions for the parameters). $B$ and $\gamma$ are in J/m$^2$ and $A$ is in GPa. $b_1$, $b_2$, $h_1$, and $h_2$ are in cubic lattice constant units ($a_0=0.36$ nm). $F$ identifies the driving force type.}
    \label{tab:parameter1}
    \begin{tabular}{|c||c||c|c||c||c|c|c||c|c|c|ccccccccccc}
    \Xhline{3\arrayrulewidth}
    {\bf GB} & {\bf F} & {\bf A} & {\bf B} & $\boldsymbol{2\gamma}$ & $\boldsymbol{b_1}$ & $\boldsymbol{h_1}$ & $\boldsymbol{\beta_1}$ & $\boldsymbol{b_2}$ & $\boldsymbol{h_2}$ & $\boldsymbol{\beta_2}$\\
    \Xhline{3\arrayrulewidth}
    \multirow{2}{*}{(310)} & \(\tau\)  & 13 & 0.39 & \multirow{2}{*}{1.9} & $\frac{1}{\sqrt{10}}$ & $-\frac{1}{\sqrt{10}}$ & $-1$ & $ \frac{1}{\sqrt{10}}$ & $\frac{3}{\sqrt{40}}$ & $\frac{2}{3}$ \\[5pt]
    \cline{2-4}
    \cline{6-11}
          & \(\psi\) & 58 & 0.39   & & 0 & $\frac{5}{\sqrt{40}}$ & $0$ & $-\frac{1}{\sqrt{10}}$ & $\frac{1}{\sqrt{10}}$ & $-1$\\[5pt]
    \hline
    \multirow{2}{*}{(510)} & \(\tau\) & 50 & 0.38   & \multirow{2}{*}{1.9} & $\frac{1}{\sqrt{26}}$ & $\frac{5}{\sqrt{104}}$ & $\frac{2}{5}$ & $\frac{1}{\sqrt{26}}$ & $-\frac{4}{\sqrt{26}}$ & $-\frac{1}{4}$\\[5pt]
    \cline{2-4}
    \cline{6-11}
          & \(\psi\) & 94 & 0.25  & & $\frac{1}{\sqrt{26}}$ & $\frac{5}{\sqrt{104}}$  & $\frac{2}{5}$ & 0 & $\frac{13}{\sqrt{104}}$ & 0 \\[5pt]
    \hline
    \multirow{2}{*}{(750)}  &  \(\tau\) & 36 & 0.53 & \multirow{2}{*}{1.5} & $\frac{1}{\sqrt{74}}$ & $-\frac{3}{\sqrt{74}}$ & $-\frac{1}{3}$ & $\frac{2}{\sqrt{74}}$ & $-\frac{6}{\sqrt{74}}$ & $-\frac{1}{3}$ \\[5pt]
    \cline{2-4}
    \cline{6-11}
          & \(\psi\) & 36 & 0.53  & & $-\frac{1}{\sqrt{74}}$ & $\frac{3}{\sqrt{74}}$ & $-\frac{1}{3}$ & $-\frac{2}{\sqrt{74}}$ & $\frac{6}{\sqrt{74}}$ & $-\frac{1}{3}$\\[5pt]
    \Xhline{3\arrayrulewidth}
    \end{tabular}
   \end{table}

The values of the parameters \(A\) and \(B\) are  of  similar magnitude for all three GBs and for both driving forces.
In fact, the non-linear fitting procedure employed showed many shallow minima, many of which give fits to the temperature-dependence of \(\beta\) of similar quality.
This suggests that the predictions for \(\beta\) based on Eqs. \eqref{betaMulModel} and \eqref{barrier} are robust (insensitive to which GB in a material, how the GB is driven, ...).

Further examination of Table \ref{tab:parameter1} shows while the values of \(B\) obtained by fitting the constant stress and constant chemical potential jump data show  little variation (the average error is less than 15\%), the value of  \(A\) is more sensitive to the type of driving force (average error ~65\%).
The value of \(A\) depends on dislocation core size \(r\) and \(r\) roughly scales with Burgers vector \(b\)~\cite{hirth1982theory}.
Stress-driven GB dynamics favors disconnections of large Burgers vector \(|\mathbf{b}|\) while GB migration  driven by a chemical potential jump favors disconnections of small \(|\mathbf{b}|\).
In fact, Table \ref{tab:parameter1} shows that the value of \(A\) obtained with stress-driven GBs is less than or equal to that obtained when migration is driven by a chemical potential jump - as expected based on this argument.
Since the two disconnections with the  lowest nucleation barriers for the (750) GB have the same value of \(\beta\), the effective coupling factor is insensitive to both temperature and driving force (see Table \ref{tab:parameter1} and  Figs. \ref{BTP} and \ref{BTE}).

The analytical expression for  \(A\) (following Eq.~\eqref{qMulModel}) depends on the core size  \(r\).
Inserting elastic constant data for Cu and using our fitted values of \(A\) (Table \ref{tab:parameter1}) suggest that \(10^{-4}<r<1\) nm.
This result shows that the expression for \(A\) is not unreasonable (and that this is not a good way to determine core size).
On the other hand, the  analytical expression for \(B\) (following Eq.~\eqref{qMulModel}) is simple and does not depend on disconnection mode: \(B=2\gamma\) \cite{hirth1982theory, han2018grain}.
Comparison of the values of \(B\) from the fitting and \(2\gamma\) from our atomistic (energy minimization) simulations shows that while the two are within an order of magnitude of one another, the agreement is not outstanding.
Hence, the analytical expressions for  \(A\) and  \(B\) should be viewed as order of magnitude estimates only.
Moreover, if the driving force is too large to linearize Eq.\eqref{betaMulModel}, fitting \(A\) and \(B\) using the linearized expression will not be accurate.

Table \ref{tab:parameter1} also gives the disconnection modes corresponding to the two lowest intrinsic nucleation barriers (see Eq. \eqref{barrier}) under the two types of driving forces.
Note that since the fitted values of \(A\) and \(B\) may depend on the type of driving force, so may the lowest intrinsic nucleation barrier modes.
The disconnections with the lowest intrinsic barriers \((b_1,h_1)\) are the same for both types of driving forces for the (510) and (750) GBs, but for the (310) GB the lowest intrinsic barrier mode is different for the stress and chemical potential jump driving forces.
For the second lowest intrinsic  barrier mode \((b_2,h_2)\), only the (750) GB chooses the same mode for both types of driving force.
In general, GB migration under a stress driving force favors disconnection modes with larger \(|\mathbf{b}|\) and smaller \(|h|\) than those under a chemical potential jump driving force  and {\it vice-versa}.

The  temperature dependent coupling factors for the (310), (510) and (750) GBs for stress and chemical potential jump-driven migration (Figs. \ref{BTP} and \ref{BTE}) may be directly compared in Fig.  \ref{BTP2}.
\(\beta\) for the (750) boundary is remarkably temperature independent (compared with the other GBs) and insensitive to the nature of the driving force.
The origin of both effects may be understood by reference to Table \ref{tab:parameter1}.
For this GB, the coupling factors for the two lowest disconnection nucleation barrier modes are identical (\(\beta_1=\beta_2=-1/3\)).
The fact that the lowest barrier mode \(\beta_1\) is the same for both driving force types implies that the coupling factor for both types of driving force will be identical at low temperature.
The fact that the second lowest barrier mode \(\beta_2\) is the same as  \(\beta_1\) implies that raising the temperature has little effect on \(\beta\).
In fact, the temperature at which we expect to see significant deviations of \(\beta\) from its low temperature value is determined by the difference in barriers between the lowest barrier mode and the next mode with a different value of \(\beta\).
For this GB, the next lowest barrier mode with a different value of \(\beta\) is the fourth lowest  (\(\beta_4\)).
Since this barrier is much larger than that of \(\beta_1\), the effective \(\beta\) deviates from its low temperature value only near the melting point.
Finally, since the two lowest  disconnection barriers are the same for both types of driving forces, \(\beta\) is independent of driving force type till very close to the melting point.

For the (510) GB, the coupling factors for the lowest disconnection nucleation barrier mode \(\beta_1\) are identical for both types of driving force.
Like for the (750) GB, this implies that as \(T\rightarrow0\), the effective coupling factor for both driving force types are the same.
However, as the temperature increases, Fig.  \ref{BTP2} shows that the coupling factors for the different driving force types diverge.
Table \ref{tab:parameter1}  shows that the second lowest barrier modes differ from the lowest barrier modes - this explains why \(\beta\) is temperature-dependent.
Table \ref{tab:parameter1} also shows that the second lowest barrier modes are different for different types of driving force  - this explains why the \(\beta(T)\) curves diverge at intermediate and high temperature.

\(\beta(T)\) for the (310) GB exhibits remarkable differences compared with the (750) and (510) GBs  (see  Fig.  \ref{BTP2}).
For this GB,  in the low temperature limit (\(T\rightarrow0\))  \(\beta\)  depends on the type of driving force.
The behavior as  \(T\rightarrow0\) is even more clear based on the continuous curves in Fig.  \ref{BTP2} which are from Eq. \eqref{betaMulModel}.
This may be explained by the fact the coupling factor corresponding to the lowest barrier mode \(\beta_1\) is different for the two types of driving forces (see Table \ref{tab:parameter1});
for \(T\rightarrow0\), Eq. \eqref{betaMulModel} shows that \(\beta\rightarrow\beta_1\).
Examination of the second lowest barrier modes (Table \ref{tab:parameter1}) shows that the difference between the first and second lowest barrier modes for the chemical potential jump-driving force is smaller (1) than that for the stress driving force (2.5).
This explains why the temperature dependence of \(\beta\) for stress-driven migration is stronger than that for chemical potential-driven migration.

\subsection*{Driving force magnitude}

Although in most studies of the coupling factor it is implicitly assumed that \(\beta\) is insensitive to the magnitude of the driving force, the results in Figs. \ref{BP2} and \ref{BE} indicate that this is not true.
Equation \eqref{betaMulModel} shows that \(\beta\) depends on driving force.  Therefore, expanding \(\beta\) to third order for small driving forces (i.e., \(bS\tau/k_{B}T\ll1\) or \(hS\psi/k_{B}T\ll1\)), we find that
  \begin{equation}\label{fitBetaStress}
    \beta=C^\tau\tau^{2}+\beta_0^\tau
  \end{equation}
  \begin{equation}\label{fitBetaEnergy}
    \beta=C^\psi\psi^2+\beta_0^\psi,
  \end{equation}
where \(\beta_0^x\) is the zero driving force limit of \(\beta\) for driving force of type \(x\) (the constants \(C^x\) and \(\beta_0^x\) are given explicitly in the Supplementary Information).
This suggests that the assumption that \(\beta\) is independent of the magnitude of the driving force is valid to first order in the driving force.
Examination of the MD simulation results shown in Figs. \ref{BP2} and \ref{BE} demonstrate that Eqs.~\eqref{fitBetaStress} and \eqref{fitBetaEnergy} provide an excellent fit to the data.
At low temperature \(T \ll T_c\), \(C^x\rightarrow0\), where \(T_c =(Q_2-Q_1)/k_B\) and \(Q_i\) is the  \(i^\text{th}\)  lowest disconnection nucleation barrier.

As discussed above,  stress-driven shear coupling and chemical potential jump-driven shear coupling may have different coupling factors \(\beta\) at small driving forces.
% , i.e., \(\beta_0\) in Eq. \eqref{fitBetaStress} and \eqref{fitBetaEnergy}, verified by Fig. \ref{BTP2}.
While this conclusion is general, it fails (i.e., \(\beta_0^\tau  = \beta_0^\psi\)) when (1) the temperature is low  and (2) \(\beta_0^\psi \neq 0\) (i.e., the lowest barrier mode does not correspond to a pure step).
This is consistent with all of the MD results in Fig. \ref{BTP2} given the lowest barrier modes shown in Table \ref{tab:parameter1}.

The fact that \(\beta\) depends on the magnitude of the driving force explains why \(\beta\) is different under constant stress and constant strain rate loading (see Fig. \ref{BTC}).
In the spirit of the derivation of the thermodynamic Maxwell relations~\cite{blundell2009concepts}, we can write
    \begin{equation}\label{stressstrain2}
        \frac{\partial\beta}{\partial T}\Big{|}_{\dot{D}}  =\frac{\partial\beta}{\partial T}\Big{|}_{\tau}+\frac{\partial\beta}{\partial\tau}\Big{|}_{T}\frac{\partial\tau}{\partial T}\Big{|}_{\dot{D}},
    \end{equation}
where \(\dot{D}\) is the relative displacement rate of the two grains meeting at the GB (the shear strain rate is \(\dot{D}/L_y\)  and \(L_y\) is the size of the bicrystal in the direction normal to the GB plane).
We note that \((\partial\tau/\partial T)|_{\dot{D}}\) is non-zero since the shear stress depends on \(\beta\) at fixed displacement rate (and, \(\tau\) decreases with increasing \(T\)).
Hence, \((\partial\beta/\partial T)|_{\dot{D}} \neq (\partial\beta/\partial T)|_{\tau}\) simply because \((\partial\beta/\partial \tau)|_T\) is
non-zero.

\section*{Conclusions}

While grain boundary migration often appears complex, we demonstrated that much of this complexity may be resolved by consideration of the underlying mechanisms by which GBs migrate.
We present molecular dynamics results that demonstrate that the temperature-dependence of the grain boundary shear coupling factor \(\beta\) (the quantity that relates GB migration to the relative translation of the grains) depends on whether the grain boundary is driven by differences (jumps) in chemical potential across the GB, stress, or strain rate.
\(\beta\) is also observed to be a function of the magnitude of the driving force.
These variations in \(\beta\) can be very large (orders of magnitude) and even lead to changes in sign.

We propose a simple model that  quantitatively predicts  these variations.
Our model is based on the statistical mechanics of disconnection (line defects in the GBs characterized by a Burgers vector and step height) nucleation.
After all crystallographically-permissible disconnection modes are predicted for any specific GB, our disconnection nucleation model determines the relative nucleation barriers for each and statistical mechanics determines the relative rates of formation of each.
We apply this theoretical construct to the four GBs examined in our MD simulations and predict which disconnections are most important for each GB and type of driving force (as well as their relative importance).
With this information, we directly predict the shear coupling factor as a function of temperature, driving force (type and magnitude), and bicrystallography.
These predictions are in excellent quantitative agreement with all of the MD simulation results.

Although the present work explicitly focused on GB dynamics in bicrystals, shear coupling constrained by multiple grains (and triple junctions) is a key feature of microstructure evolution in polycrystals (including grain size coarsening, grain rotation, stress generation, etc \cite{thomas2017reconciling}).  Disconnections of one mode, moving along GBs in polycrystals, will pile-up at triple junctions, creating back stresses that will prevent macroscopic GB migration. Continued GB (and triple junction) migration requires the participation of disconnections of other modes to ensure that the total Burgers vector is zero at the triple junction~\cite{han2018grain}. The current work presents a statistical mechanics-based approach that provides the basis for explaining how multiple disconnection modes  conspire to move GBs and triple junctions together.

\section*{Simulation Method}
We perform MD simulations of several symmetric tilt GBs in copper employing the Large-scale Atomic/Molecular Massively Parallel Simulator (LAMMPS)~\cite{plimpton1995fast} using periodic boundary conditions and an embedded atom method-type of interatomic potential fit to copper~\cite{Mishin2011}.
We apply both stress and chemical potential jump driving forces (larger driving forces were used for the \(\Sigma13(510)\) GB than the others because the mobility of this boundary is considerably).
All simulations are 7 ns in duration at temperatures in the \(600-1300\) K range.
The GB position is determined as the maximum in the centro-symmetry parameter~\cite{kelchner1998dislocation} in the direction normal to the GB plane using the visualization package  OVITO~\cite{stukowski2009visualization}.
The shear coupling factor \(\beta\)  for the two GBs in the simulation cell are measured separately from the ratio of the translation rate of the grains parallel to the GB to that of the mean GB plane normal velocity.

\bibliography{mybib}
 \end{document}

\title{SUPPLEMENTARY INFORMATION \\ \vspace*{0.3cm} for \\ \vspace*{0.3cm} Grain Boundary Shear Coupling  is Not a Grain Boundary Property}
   \author{Kongtao Chen$^1$}
   \author{Jian Han$^1$}
   \author{Spencer L. Thomas$^1$}
   \author{David J. Srolovitz$^{1,2,3}$}

\affiliation{$^1$Department of Materials Science and Engineering, University of Pennsylvania, Philadelphia, Pennsylvania 19104, USA}
\affiliation{$^2$Department of Mechanical Engineering and Applied Mechanics, University of Pennsylvania, Philadelphia, Pennsylvania 19104, USA}
\affiliation{$^3$Department of Materials Science and Engineering, City University of Hong Kong, Hong Kong SAR, PRC}

   \date{\today}

   \begin{abstract}
    \end{abstract}

   \maketitle\section{Simulation Details}

We construct GBs by fixing the misorientation of the two crystals and the GB plane and minimize the energy with respect to atomic coordinates and the relative translations of the upper grain relative to the lower grain parallel to the GB plane.
All of our periodic bicrystals are $\sim25$ nm in the $x-$direction,  $100$ nm in the $y-$direction, and three lattice constants (1.1 nm) in the $z-$direction.
We then rescale all atomic coordinates in accordance with temperature-dependent lattice constant prior to simulations at any temperature.
Before applying a driving force, we equilibrate the bicrystal system at the temperature of interest for 0.2 ns.
When applying constant strain rate \(\dot{\epsilon}_{xy}\), we shear the simulation cell at a constant rate while [\onlinecite{Cahn2006}] apply constant velocities to atoms in the top and bottom regime of the simulation cell.

\section{Analytical Expression for \(C^x\) and \(\beta_0^x\)}
We can expand the shear coupling factor \(\beta\) for stress-driven GB migration (Eq. (2) in the main text)
\begin{equation}\label{betaMulMode}
\beta=\frac{\sum_{i}b_{i}e^{-\frac{Q_{i}}{k_{B}T}}\sinh{\frac{(b_{i}\tau-h_{i}\psi)S}{k_{B}T}}}{\sum_{i}h_{i}e^{-\frac{Q_{i}}{k_{B}T}}\sinh{\frac{(b_{i}\tau-h_{i}\psi)S}{k_{B}T}}}
\end{equation}
at high temperature to third order in \(bS\tau/k_{B}T\). This yields $\beta=C^\tau\tau^{2}+\beta_0^\tau$, where

\begin{equation}\label{fitBetaStress1}
      C^\tau=\frac{\frac{S^2}{6k_B^2T^2}(\sum_ie^{-\frac{Q_{i}}{k_{B}T}}b_i^4-\frac{\sum_ie^{-\frac{Q_{i}}{k_{B}T}}b_i^2\sum_ie^{-\frac{Q_{i}}{k_{B}T}}h_ib_i^3}{\sum_ie^{-\frac{Q_{i}}{k_{B}T}}h_ib_i})}{\sum_ie^{-\frac{Q_{i}}{k_{B}T}}h_ib_i}
\end{equation}
\begin{equation}\label{fitBetaStress2}
      \beta_0^\tau=\frac{\sum_ie^{-\frac{Q_{i}}{k_{B}T}}b_i^2}{\sum_ie^{-\frac{Q_{i}}{k_{B}T}}h_ib_i}.
\end{equation}

Similarly, expanding Eq.~\eqref{betaMulMode} at high temperature to third order in  \(hS\psi/k_{B}T\)  yields \(\beta=C^\psi\psi^{2}+\beta_0^\psi\), where

\begin{equation}\label{fitBetaStress1}
      C^\psi=\frac{\frac{S^2}{6k_B^2T^2}(\sum_ie^{-\frac{Q_{i}}{k_{B}T}}h_i^3b_i-\frac{\sum_ie^{-\frac{Q_{i}}{k_{B}T}}h_ib_i\sum_ie^{-\frac{Q_{i}}{k_{B}T}}h_i^4}{\sum_ie^{-\frac{Q_{i}}{k_{B}T}}h_i^2})}{\sum_ie^{-\frac{Q_{i}}{k_{B}T}}h_i^2}
\end{equation}
\begin{equation}\label{fitBetaStress2}
      \beta_0^\psi=\frac{\sum_ie^{-\frac{Q_{i}}{k_{B}T}}h_ib_i}{\sum_ie^{-\frac{Q_{i}}{k_{B}T}}h_i^2}.
\end{equation}

At relatively low temperature (\(b\tau S\ll k_{B}T\ll Q\), \(h\psi S\ll k_{B}T\)), \(C^x\rightarrow0\).
This implies that at low temperature, the shear-coupling factor is determined by the lowest barrier mode and is insensitive to magnitude of driving force.

\section{Limiting Behavior}
The driving force affects the shear-coupling factor \(\beta\) by tilting the energy landscape thereby lowering  disconnection nucleation barriers.
While large stresses favor disconnection modes of (relatively) large Burgers vector and small step height, large chemical potential jumps favor modes of (relatively) small Burgers vector and large step height (especially pure step modes where \(\mathbf{b}=0\)).
This implies different values of \(\beta\) for different driving forces.
More specifically, under an applied stress, \(|\beta|\) tends to be larger than under a chemical potential jump.
Even in the small driving force limit, \(\beta\) depends on the type of driving force; except in special cases, e.g., at low temperature and finite \(\beta\).
When the stress-driving force is large (\(\tau\to\infty\)), GB sliding occurs (\(\beta\to\infty\)); this is because stress favors modes of large Burgers vector and negligible step height.
On the other hand, when the chemical potential jump driving force is large (\(\psi\to\infty\)), the GB migrates without coupling (\(\beta\to0\)); this is because chemical potential jumps favor  modes of negligible Burgers vector and large step height.

Even at high temperatures, GBs migrate by the nucleation and motion of disconnections~\cite{han2018grain}, each with a specific Burgers vector (leading to grain translation) and step height (leading to GB migration).
In the high temperature limit (\(T\to\infty\)), many disconnections modes are activated simultaneously.
Hence, chemical potential jump-driven GB migration occurs with no net grain translation (\(\beta\to0\)) because disconnections with the same sign of \(h\) but opposite signs of \(\mathbf{b}\) are nucleated such that the average \(\mathbf{b}=0\).
Similarly, when driven by stress at high temperature, grain boundaries slide (\(\beta\to\infty\)) because disconnections with the same sign of \(\mathbf{b}=0\) but opposite signs of \(h\) are nucleated such that the average \(h=0\).

\bibliography{mybib}